\documentclass[journal=jacsat,manuscript=article]{achemso}

\usepackage[english]{babel}
\usepackage[utf8]{inputenc}
\usepackage{xr-hyper}
\usepackage{array}

\usepackage[colorlinks]{hyperref}
\usepackage[T1]{fontenc}
\usepackage{amsmath}
\usepackage{amsfonts}
\usepackage[dvipsnames]{xcolor}
\usepackage[caption=false, justification=justified]{subfig}

\usepackage{url}
\usepackage{ulem}
\usepackage{braket}

\usepackage{multirow}

\newcommand{\ang}{Å} 

\hypersetup{filecolor=black}
\hypersetup{citecolor=black}
\hypersetup{urlcolor=black}
\hypersetup{linkcolor=black}

\title{Two-dimensional to bulk crossover of the WSe\textsubscript{2} electronic band structure}
\date{\today}
\author{Raphaël~Salazar}
\affiliation{Synchrotron SOLEIL, L’Orme des Merisiers, Départementale 128, F-91190 Saint-Aubin, France}
\alsoaffiliation{New Technologies Research Centre, University of West Bohemia, 30614 Pilsen, Czech Republic}
\author{Matthieu~Jamet}
\affiliation{Univ. Grenoble Alpes, CEA, CNRS, Grenoble INP, IRIG-SPINTEC, 38000 Grenoble, France}
\author{Céline~Vergnaud}
\affiliation{Univ. Grenoble Alpes, CEA, CNRS, Grenoble INP, IRIG-SPINTEC, 38000 Grenoble, France}
\author{Aki~Pulkkinen}
\affiliation{New Technologies Research Centre, University of West Bohemia, 30614 Pilsen, Czech Republic}
\author{François~Bertran}
\affiliation{Synchrotron SOLEIL, L’Orme des Merisiers, Départementale 128, F-91190 Saint-Aubin, France}
\author{Chiara~Bigi}
\affiliation{Synchrotron SOLEIL, L’Orme des Merisiers, Départementale 128, F-91190 Saint-Aubin, France}
\author{J\'an~Min\'ar}
\affiliation{New Technologies Research Centre, University of West Bohemia, 30614 Pilsen, Czech Republic}
\author{Abdelkarim~Ouerghi}
\affiliation{Université Paris-Saclay, CNRS, Centre de Nanosciences et de Nanotechnologies, 91120, Palaiseau, Paris,
France}
\author{Thomas~Jaouen}
\affiliation{Univ Rennes, CNRS, IPR - UMR 6251, F-35000 Rennes, France}
\author{Julien~Rault}
\affiliation{Synchrotron SOLEIL, L’Orme des Merisiers, Départementale 128, F-91190 Saint-Aubin, France}
\affiliation{ABB Switzerland Ltd, Baden Dättwil, Switzerland}
\author{Patrick~Le Fèvre}
\email{patrick.lefevre@univ-rennes.fr}
\affiliation{Synchrotron SOLEIL, L’Orme des Merisiers, Départementale 128, F-91190 Saint-Aubin, France}
\alsoaffiliation{Univ Rennes, CNRS, IPR - UMR 6251, F-35000 Rennes, France}

\keywords{Transition metal dichalcogenides, 2D-materials, ARPES, molecular beam epitaxy, tight-binding calculation, KKR-Green's function method}

\begin{document}

\begin{abstract}
Transition Metal Dichalcogenides (TMD) are layered materials obtained by stacking two-dimensional sheets weakly bonded by van der Waals interactions. In bulk TMD, band dispersions are observed in the direction normal to the sheet plane ($z$-direction) due to the hybridization of out-of-plane orbitals but no $k_z$-dispersion is expected at the single-layer limit. Using angle-resolved photoemission spectroscopy, we precisely address the two-dimensional to three-dimensional crossover of the electronic band structure of epitaxial WSe$_2$ thin films. Increasing number of discrete electronic states appears in given $k_z$-ranges while increasing the number of layers. The continuous bulk dispersion is nearly retrieved for 6-sheet films. These results are reproduced by calculations going from a relatively simple tight-binding model to a sophisticated KKR-Green's function calculation. This two-dimensional system is hence used as a benchmark to compare different theoretical approaches.
\end{abstract}

\begin{figure}[h!]
    \centering
    \includegraphics[width=16cm] {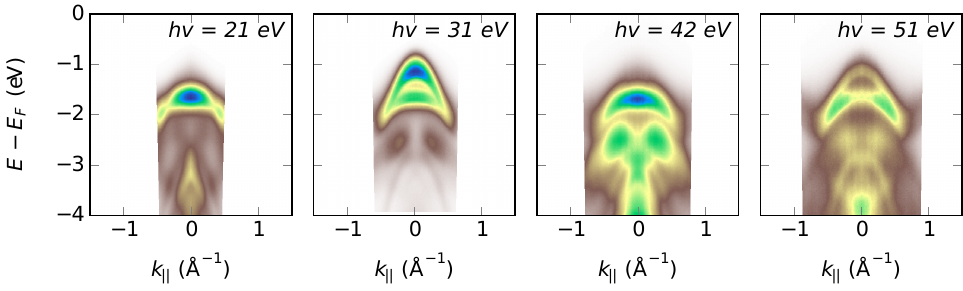}
    \caption{ARPES on 2-layer WSe$_2$ giving the band-dispersion in directions parallel to $\Gamma-K$ at various photon energies (21, 31, 42 and 51 eV, from left to right). Depending on the photon energy, one or two bands are visible at $\Gamma$ between -2 eV binding energy and the Fermi level. Spectra have been symmetrized with respect to $\Gamma$.}
    \label{fig:2ML}%
\end{figure}

Ever since the discovery of graphene \cite{Novoselov2004,Novoselov2005}, research on two-dimensional (2D) materials is ongoing a tremendous effort. Along with this trend, transition metal dichalcogenides (TMD) are extremely promising for possible technological applications. Their general formula is MX$_2$, where M is a transition element and X is a chalcogene. Those two elements form a MX$_2$ basic layer, where M atoms are sandwiched between two covalently-bonded planes of chalcogenes. The three-dimensional (3D)-solid is obtained by stacking these X-M-X sheets, only weakly bonded by van der Waals interactions \cite{Dickinson1923}, conferring to TMD a very pronounced 2D-character. Nowadays, research pushes towards increasingly elaborated structures taking advantage of the 2D nature of these materials: twisted TMD layers \cite{Gatti2023}, hybrid TMD structures (e.g., MoSe$_2$/WSe$_2$, WS$_2$/WSe$_2$) \cite{Yuan2020, Stansbury2021, Khalil2022} or alloyed TMD systems \cite{Ernandes2021}. In the ideal case, these structures are studied by the means of Angle-Resolved PhotoEmission Spectroscopy (ARPES), a technique which allows for direct measurement of the band structure. Its surface sensitivity makes it particularly well-suited to probe 2D-compounds. For bulk TMD-crystals, despite the expected strong 2D-nature of these materials, a significant band dispersion can be observed perpendicular to the MX$_2$-sheets \cite{Finteis1997, Riley2014, Kim2016}. As a matter of fact, both the M-$d_{z^2}$ and the X-$p_z$ orbitals (the $z$-axis being perpendicular to the MX$_2$-sheets) point out from the MX$_2$-planes and can hybridize to give rise to this perpendicular dispersion. It can be successfully modeled either using Density Functional Theory (DFT) \cite{Riley2014} or a tight-binding approach \cite{Kim2016}. Nevertheless, DFT-band structure calculations are often made assuming the 2D character of TMD and obtain results on a mesh with only one $k_z$ point. For instance, from calculations presented in reference \citenum{Nguyen2019}, one could hastily conclude that, at low thicknesses, each new layer in the stacking generates an additional band at $\Gamma$. Figure \ref{fig:2ML} shows dispersions measured on a 2-layer WSe$_2$ sample in directions parallel to $\Gamma-K$ for various photon energies, \textit{i.e.} various position along the $\Gamma-A$ direction. We do observe the expected two bands at 31 and 51 eV-photon energies, but not at 21 and 42 eV. That means that even for ultimately thin samples, $k_z$-effects can be expected, which is more surprising. Previous work on single and bi-layer MoS$_2$ already hinted at those conclusions without proposing a complete explanation \cite{Miwa2015}. These primary observations urge towards a better understanding of the $k_z$-dependency even for thin samples.

We propose here to study the transition from the 2D-electronic structure of an ultimately thin TMD to a 3D-bulk crystal. We focused on WSe$_2$, whose electronic structure has already been extensively scrutinized \cite{Riley2014,Kim2016,Nguyen2019}. The perpendicular dispersion of the band structure is measured by ARPES for sample thicknesses of 2, 3 and N (=6--7) layers. The 2- and N-layer samples were grown by molecular beam epitaxy (MBE) on graphene/SiC (Gr-SiC). To complete the experiment, a 3-layer sample of WSe$_2$ was grown on Mica and then wet-transferred onto a Gr-SiC substrate \cite{Salazar2022,Dau2019}. More details about the samples are given in the Supporting Information.

The ARPES measurements were performed at room temperature on the CASSIOPEE beamline of the SOLEIL synchrotron radiation facility. The samples were first aligned with the $\Gamma$-K direction of WSe$_2$ reciprocal lattice along the slit of the analyzer. We call $k_{\parallel}$ the component of the wave vector parallel to this direction. A ($k_{\parallel}$, $E_B$) image can then be measured at once thanks to the 2D-detector of the electron analyzer. Here, $E_B$ is the electron binding energy and is measured with respect to the Fermi level $E_F$. $k_z$ was changed by scanning the photon energy from 20 to 90 eV (and 1 eV-step), which amounts to span a $k_z$-range from roughly 2.5 to 5 \ang$^{-1}$. Details on the experimental geometry, on the measurement strategy and on the determination of $E_F$ are given in the Supporting Information.

\begin{figure}[h!]
    \centering
    \includegraphics[width=16cm]{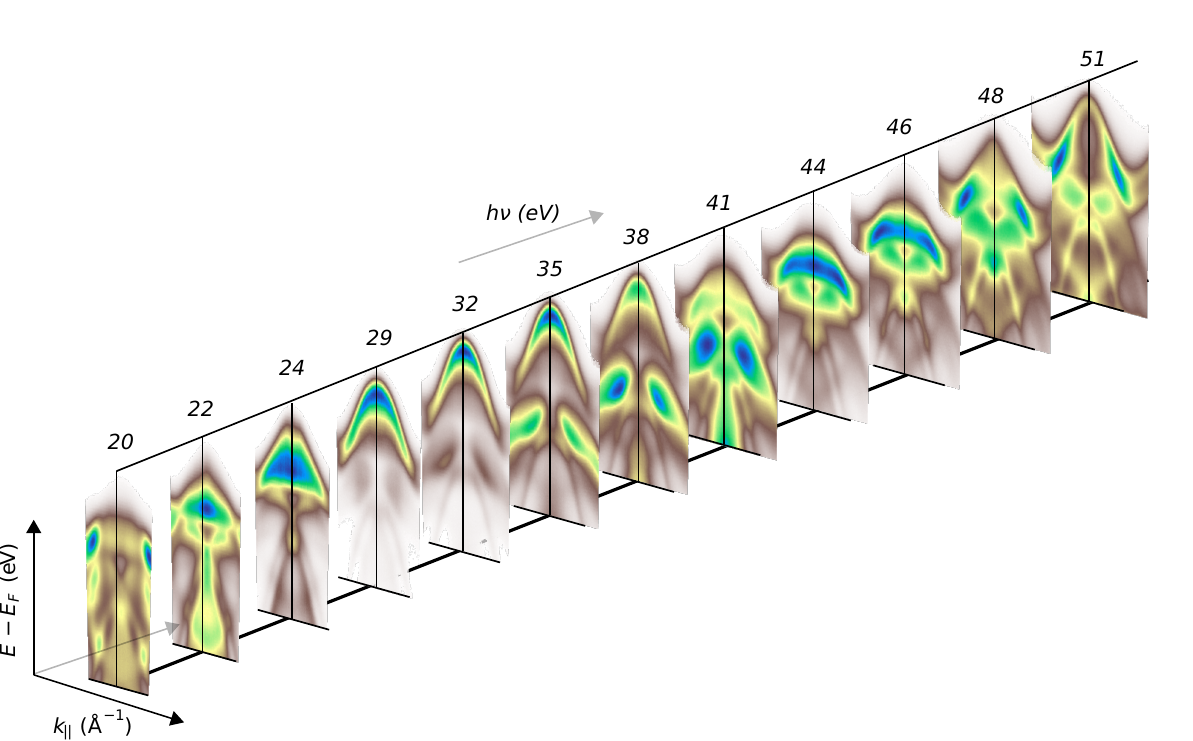}
    \caption{Series of ARPES-images recorded on the N-layer (N=6-7) WSe$_2$/Gr-SiC sample showing the dispersion along $k_{\parallel}$ for chosen photon energies between 20 and 90 eV. The vertical black line is at $k_{\parallel}$=0.}
    \label{fig:ImageSeries}%
\end{figure}

\begin{figure}[h!]
    \centering
    \includegraphics[width=16cm]{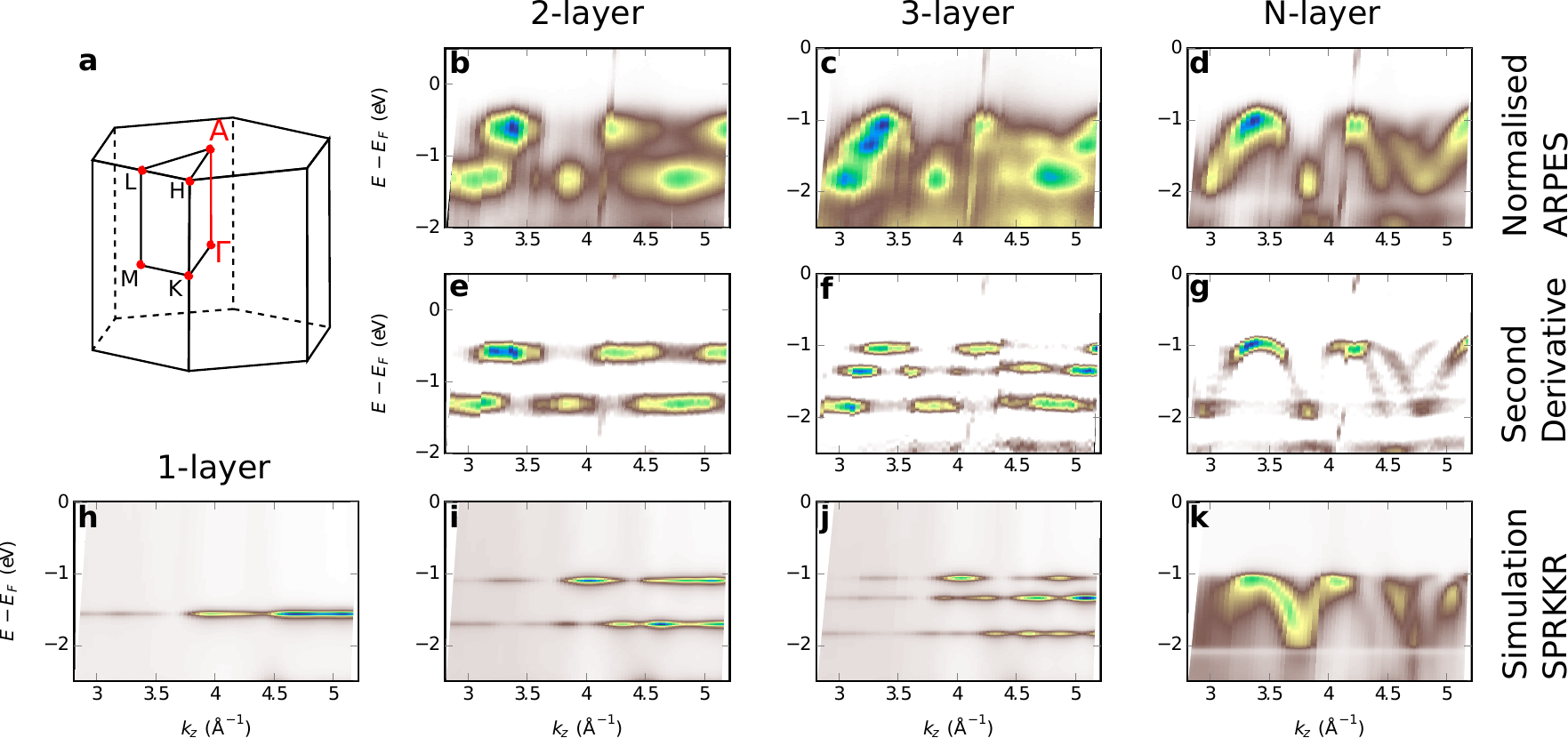}
    \caption{(a) First Brillouin zone of the WSe$_2$ reciprocal space. (b-g) Experimental band intensity variations at $\Gamma$ along the $\Gamma-A$ direction of the reciprocal space for 2 (left), 3 (middle) and N (right) layers of WSe$_2$. Panels (b-d) show the raw data and panels (e-g) their second derivative. Panels (i-k) show the corresponding calculations done in the KKR-Green's function formalism (see text for details) with an additional calculation for 1-layer WSe$_2$ (h). The $h\nu$ to $k_z$ conversion was done using $V_0 =$ 13 eV for all spectra (See Supporting Information).}
    \label{fig:kzdispersion}%
\end{figure}

Figure \ref{fig:ImageSeries} shows a series of ARPES-images recorded on the N-layer sample showing the dispersion along $k_{\parallel}$ for chosen photons energies between 20 and 90 eV. It clearly evidences large variations of the band structure along $\Gamma-A$ (see Figure \ref{fig:kzdispersion}(a)). The vertical black line is at $k_{\parallel}$=0.

To have a clear view of the $k_z$-dispersion along $\Gamma-A$, we plot this cut at $k_{\parallel}$=0 as a function of the perpendicular component of the wave vector on Figure \ref{fig:kzdispersion} for the different samples (same data as a function of the photon energy are available in the Supporting Information). Figures \ref{fig:kzdispersion}(b-d) show the raw data for 2, 3 and N-layers samples. For the 2-layer sample, the two bands which are the closest to the Fermi level clearly have a discontinuous $k_z$-dispersion. It consists in line segments corresponding to $k_z$-values where these bands appear at $k_{\parallel}$=0 in the $\Gamma-K$ dispersion. Hence, the top band is visible only for $k_z$ between 3.1 and 3.5 \ang$^{-1}$ (photon energies between 23 and 40 eV); it then disappears to show up again between 4 and 4.6 \ang$^{-1}$ (50 and 70 eV). It has a low intensity from 4.5 to 5 \ang$^{-1}$ (70 to 80 eV) and becomes bright again from 5 \ang$^{-1}$ onwards (85 to 90 eV). It therefore appears and disappears when varying the photon energy, confirming the first observations made on Figure \ref{fig:2ML}. Near, e.g., $k_z$ = 3 or 3.7 \ang$^{-1}$ (20 eV and 43 eV) applying the rule 1 layer = one band at $\Gamma$, the sample is indistinguishable from a monolayer system with only one visible band at $\Gamma$. Let us note that this is the case around 21 eV, the photon energy produced by He-lamps. The same phenomenon is observable on the 3-layer sample with instead three bands appearing and disappearing. We note that the binding energies are slightly shifted in the 3-layer sample with respect to the 2-layer sample. This derives from the different thicknesses of the Gr-SiC substrate which induce differentiated charge transfers \cite{Zhang2021}. The reader can get more details in the Supporting Information. Nevertheless, the bright segments are shorter in photon energy and more numerous for each band. It starts to draw the clear band oscillations that we clearly see in the N-layer case, where the observed dispersion compares well with previously measured data on bulk-WSe$_2$ \cite{Riley2014,Kim2016}. The same observations can be made on the second derivative images (Figure \ref{fig:kzdispersion}(e-g)).

At these very small thicknesses, the electronic states considered here can certainly be seen as 2D-electronic systems, confined within the few atomic layers forming the WSe$_2$-layer. They can then be described as quantum well states whose behaviour in ARPES has been extensively described. The pioneering work of Louie \textit{et al.} \cite{Louie1980} was the first to evidence intensity variations of the Cu(111) Schockley surface state when varying the photon energy, with maxima for $k_z$-final states corresponding to the $L$-point of the 3D Brillouin zone. Intensity resonances of surface states using photon energies matching vertical transitions at high symmetry points of the 3D Brillouin zones was later confirmed by, e.g., studies on the Al(001) \cite{Hofmann2002} or Cu-vicinal surfaces \cite{Lobo2006}. This is precisely what we observe here, although with some complications coming from the WSe$_2$ crystallographic structure. Taking a $c$-parameter of 12.96 \ang~\cite{Schutte1987}, one obtains a $\Gamma-A$ distance of 0.2424 \ang$^{-1}$. At low thicknesses (2 and 3-layer samples), the maxima observed in our data (around 3.15, 4.12 and 5.1 \ang$^{-1}$) correspond to the positions of the $A_6$, $A_8$ and $A_{10}$-points of the reciprocal space. These maxima are observable on all the samples. The periodicity appears therefore as doubled in the reciprocal space, as compared to what is expected. This has been explained by Finteis \textit{et al.} \cite{Finteis1997}, following the works of Pescia \textit{et al.} \cite{Pescia1985}. WSe$_2$ has a hexagonal structure whose primitive cell contains two WSe$_2$-layers (2H-WSe$_2$). It belongs to the $D^4_{6h}$-space group which is nonsymmorphic, \textit{i.e.}, it includes a screw axis which is located in the center of the unit cell along the $c$-axis. Group theory implies selection rules for photoemission from the Bloch states on the $\Gamma-A$ line of the Brillouin zone that are restricted to the subgroups of $\Delta_1$ and $\Delta_2$ symmetry, respectively, which together with the even parity required for coupling to a free-electron final state, results to allowed optical transition from a given initial state every other Brillouin zone \cite{Finteis1997}. 

Our results look very much alike what was obtained on graphene by Ohta \textit{et al.} \cite{Ohta2007}. On a single layer sample, they showed that the $\pi$-orbital (forming the famous Dirac cone at the $K$-point of the reciprocal space) is confined in the crystal plane and show no $k_z$-dispersion. As the number of graphene layers increases, the number of $\pi$-orbital increases because of interlayer interactions and more and more discrete states appear, gradually converging towards the final 3D-dispersion \cite{Ohta2007}. This is clearly the model system to which our data on WSe$_2$ should be compared. An apparent doubling of the reciprocal space periodicity is also observed, since graphite structure also belongs to a nonsymmorphic space group \cite{Pescia1985}. The interpretation of these results was nicely revisited by Strocov in 2018 \cite{Strocov2018}. In this work, ARPES is first interpreted in a Fourier-transform formalism and quantum confined 2D-states are introduced as standing waves multiplied by an envelope function quickly decreasing away from the surface. This lucid model not only confirms that intensity maxima should appear in ARPES at high symmetry points along $k_z$, but also predicts that these maxima spread over a $k_z$-range inversely proportional to the $z$-spatial extension of the 2D-state \cite{Strocov2018}. In other words, the "dots" observed at fixed binding energies should gradually shorten as the $z$-delocalization increases. This is what is observed on graphene \cite{Ohta2007} as well as in our results. Finally, let us note that this intensity behaviour is also well-reproduced by tight-binding calculations performed on a one-dimensional atomic chain of increasing length \cite{Moser2017}. In the following, we'll show that different theoretical models can account for this behaviour, including fine details.

To understand better its physics, we propose to study the system at different levels of approximation using first a tight-binding initial state and free electron (FE) final state (TB-FE model). We then increase the complexity by describing the photoemission process within the one-step model of photoemission~\cite{BME18} as implemented in the SPRKKR package~\cite{Ebert2011} using a free electron final state (1-step-FE model). The last step is the calculation using the one-step model and a time-reversed LEED (TrLEED) final state (1-step-TrLEED model). This last result is presented in Figure \ref{fig:kzdispersion}(h-k) for free-standing 1, 2, 3 and an infinite number of WSe$_2$ layers. They are in excellent agreement with the measured data, reproducing the discontinuous patterns and converging to a bulk-like dispersion. 

\begin{figure}[h!]
    \centering
    \includegraphics[width=16cm]{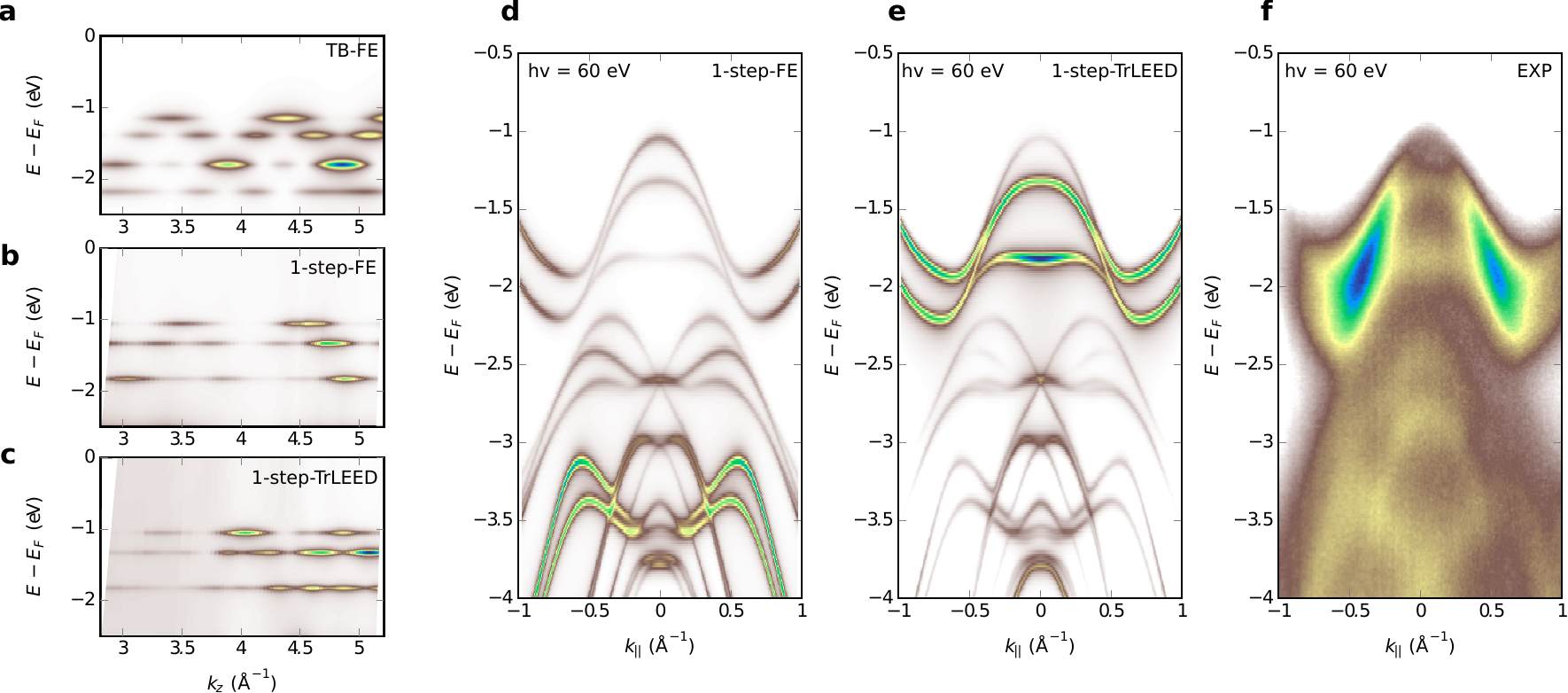}
    \caption{$k_z$-dispersions for the 3-layer system (a) from tight-binding initial state and free electron final state model (TB-FE), (b) from Bloch spectral function initial state and free electron final state (1-step-FE), (c) from Bloch spectral function initial state and time-reversed LEED final state (1-step-TrLEED). (c) is normalised to the background intensity. 3-layer sample ARPES dispersion at $\Gamma$ for $h\nu = 60$ (d) calculated in the 1-step-FE model, (e) from the 1-step-TrLEED model and (f) as measured by ARPES. No normalization of intensities have been applied for (d-f).}
    \label{fig:model_comp}%
\end{figure}

The tight-binding model \cite{GitlabRaph} is inspired from the derivation made in references \citenum{Kim2016} and \citenum{amorim2018}. We aim at a minimal model valid at $\Gamma$ along the $k_z$ direction. For more complete derivations, the reader can refer to references \citenum{Cappelluti2013,Roldan2014,Fang2015,SilvaGullen2016,amorim2018}. More details about our tight-binding model and the calculation of the photoemission current can be found in the Supporting Information. Figure \ref{fig:model_comp} shows the results of the three types of calculation on the trilayer system. In Figure \ref{fig:model_comp}(a), we see that the simple tight-binding model already captures the essential characteristics of the system with discrete energy states appearing in $k_z$-ranges in a staggered fashion. It underestimates the total amplitude of the dispersion (difference between the lowest level and highest energy level) of 0.2 eV (See Supporting Information for a quantitative comparison of these energy differences at $\Gamma$ and $K$). The intensities are only indicative and the model does not resolve the complex symmetry effects due to the screw-axis \cite{Finteis1997}. For this reason, we only show the contribution of the phase corresponding to the dominant photoemission intensity in experiments ($\varphi = 2\pi/c$, $c/2$ being the interlayer distance, see Supporting Information). Figure \ref{fig:model_comp}(b) shows the results for the 1-step-FE model. They are strikingly similar to those of the TB-FE model. They nevertheless show a better agreement with experimental data: the values of the energy levels are more precisely calculated (with discrepancies smaller than 20 meV) and the photoemitted intensity displays additional modulation along $k_z$. The symmetry effects due to the crystal space group are now consistent with the experimental data. The pattern matches especially well with the measurements in the 2.5 to 4 \ang$^{-1}$ $k_z$-range. Finally, Figure \ref{fig:model_comp}(c) shows the outcome of the 1-step-TrLEED model. In essence, the modulation of intensity is a step closer to experimental data. The $k_z$-patterning matches extremely well the experiment in the 3.5 to 5.5 \ang$^{-1}$ range. On the other hand, we notice a discrepancy with an excess intensity in the 4.5 \ang$^{-1}$ and onwards range for the top band at -1 eV. Comparing ARPES dispersions in Figure \ref{fig:model_comp}(d,e),  we see that the 1-step-FE model grossly overestimates the spectral weight of lower energy bands comparatively to those at the top of the valence bands. In the 1-step-TrLEED, the distribution of the spectral weight is improved: the top of the valence bands generally has a higher intensity than the lower lying bands in the experimental data (see Figure \ref{fig:model_comp}(f)). 

To conclude, the ARPES study of MBE-deposited WSe$_2$ films with variable thickness gives an overview of the evolution of the electronic structure of this TMD during its transition from 2D to 3D. The behaviour observed at thicknesses as low as 2 or 3-layer WSe$_2$, with discrete states appearing at constant binding energies over finite $k_z$-ranges, are coherent with the predicted signature of 2D-states, confined in the plane of the atomically-thin crystal. Their evolution with an increasing number of layers shows a larger and larger delocalisation as interlayer electronic hoppings become possible. A 6-7 layer film already shows an electronic structure comparable to what was measured on bulk crystals. Phenomenological \cite{Strocov2018} or simplified \cite{Moser2017} models, as well as what is usually observed in 2D surface states account well for our observations, highly similar to what was previously measured on graphene layers \cite{Ohta2007}. Here, the results were completely modeled by various methods with increasing complexity, as a bench test for these models: a simple tight-binding model accounts for most of the experimental observations but does not capture the effects of the crystal symmetry on the photoemission signal. An \textit{ab initio} calculation in the KKR-Green's function formalism using a free electron final state is more accurate but fails to reproduce the observed relative intensities of the different bands. A similar calculation using a TrLEED final state improves a lot this aspect. These last two calculations are performed on systems with the real geometry, going from 2D to 3D. The electronic properties of TMD, e.g., the nature (indirect or direct) of their band gap, strongly vary with thickness, both in "classical" TMD \cite{Mak2010,Radisavljevic2011}, or in more sophisticated but close compounds \cite{Ernandes2021}. It is therefore of prime interest to know how the electronic structure evolves, as described by our results.

\section{Associated content}
Additional experimental details, materials, and methods

\section{Acknowledgments}
R.S. acknowledges the support of the French National Research Agency (ANR) (CORNFLAKE project, ANR-18-CE24-0015-01). This work was supported by the project Quantum materials for applications in sustainable technologies (QM4ST), funded as project No. CZ.02.01.01/00/22$\_$008/0004572 by Programme Johannes Amos Commenius, call Excellent Research. This publication was supported by the project TWISTnSHINE, funded as project No. LL2314 by Programme ERC CZ.

\bibliography{References.bib}

\end{document}


\newpage
\section{ARPES measurements}
\begin{figure}[!h]
    \centering
    \includegraphics[scale = 2]{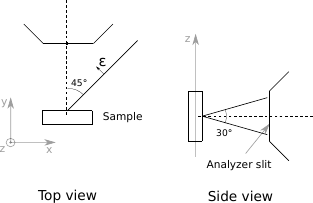}
    \caption{Top and side view of the experimental geometry showing the sample (schematized as a rectangle) and the nose of the electron analyzer. On the top view, the light is coming from the right (schematized as a solid line) at 45\deg~ from the sample normal when it is facing the analyzer. $\mathbf{\varepsilon}$ is the polarisation vector of the beam, which was always linear horizontal in our measurements. The slit of the analyzer is vertical (along the $z$-axis, as defined at the bottom left of the figure).}
    \label{fig:analyzer_geometry}
\end{figure}

The ARPES measurements were all performed at the CASSIOPEE beamline of the SOLEIL storage ring using a Scienta R4000 analyzer. The sample is installed on a vertical sample holder, with its surface normal horizontal. The photon beam is horizontal and comes at 45\deg~ from the electron analyzer axis (see Figure \ref{fig:analyzer_geometry}), which is mounted with its entrance slit vertical. In our measurements, we used linear horizontal (LH) polarized light. Three rotations are available to precisely align the sample with respect to the electron analyzer.

\begin{itemize}
    \item A $\theta$-rotation around the $z$-vertical axis (see bottom left of Figure \ref{fig:analyzer_geometry} for the definition of the frame). This rotation is used to perform a complete 3D-band structure measurement ($k_x$,$k_y$,$E_B$) which can be cut at any electron binding energy $E_B$, giving the constant energy surfaces used below to align and characterize the samples (see Figure \ref{fig:3D_N_ML}).
    
    \begin{figure}[!h]
    \centering
    \includegraphics[scale = 1.3]{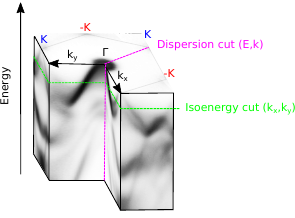}
    \caption{Example of a 3D data block ($k_x$,$k_y$,$E_B$) of the band structure of the N-layer sample.}
    \label{fig:3D_N_ML}
\end{figure}

     \item A $\phi$-rotation around the sample surface normal, which can be used to align any crystallographic axis of the sample along the analyzer slit.
    \item A Tilt-rotation around a horizontal axis contained in the sample surface plane, which was mainly used here to correct vertical angular misalignment.
\end{itemize}

For the photon energy dependant measurements presented in the main text of the paper, we used these three degrees of freedom to align the $\Gamma$-$K$ direction of the sample reciprocal space along the electron analyzer slit. Three translations along perpendicular axis are also available, first to place the sample surface at a correct measurement position, but also possibly to scan the sample surface. In this latter case, the spatial resolution is around 100 µm. All the measurement were performed at room temperature. We call $k_{\parallel}$ the component of the wave vector parallel to this direction. A ($k_{\parallel}$, $E_B$) image can then be measured at once thanks to the 2D-detector of the electron analyzer. Here, $E_B$ is the electron binding energy and is measured with respect to the Fermi level. The most efficient way to travel along $k_z$ in the reciprocal space ($\Gamma-A$ direction) is to vary the photoelectron kinetic energy $E_K$. For a photoelectron going out from the sample along the surface normal, the relation between $k_z$ and $E_K$ is given by $k_z=\sqrt{\frac{2m}{\hbar^2}(E_K+V_0)}$, where $m$ is the electron mass and $V_0$ is the so-called inner potential, a material-dependent quantity which is not known \textit{a priori} but can be determined experimentally \cite{Hufner2003}. In bulk WSe$_2$, it was shown to be in the order of 13 eV \cite{Riley2014} (14.5 eV in the work of Finteis \textit{et al.} \cite{Finteis1997}). In this work, we kept the value $V_0 = 13$ eV for all conversions. Since $E_K\simeq h\nu - E_B$, varying the photon energy corresponds to changing $k_z$. In our measurements, we used photons from 20 to 90 eV (and 1 eV-step), which amounts to span a $k_z$-range from roughly 2.5 to 5 \ang$^{-1}$.

The principle of the monochromator installed on CASSIOPEE does not allow for an absolute determination of the photon energy. To calculate as precisely as possible the relative binding energies from one measurement to the other, the Fermi level energy was measured every 5 eV (at 20, 25... and 90 eV photon energies) on the Mo-clips holding the sample and connecting it to the ground. The Fermi level energy position was then extracted for each photon energy by linearly interpolating the data set, and used as the reference for the binding energies. To allow for more quantitative analysis, the intensities of the spectra were normalised to the secondary electron background intensity above the Fermi level (excited by higher harmonics of the undulator providing photons to the beamline), correcting both the detector background and the differences in flux between two photon energies.

\section{More details about the samples}
Most of the samples were grown by Moleclar Beam Epitaxy (MBE) on graphene/SiC(0001), held at 573 K as measured by a thermocouple in contact with the sample holder) by co-evaporating W from an e-gun evaporator at a rate of 0.15 \ang/min and Se from an effusion cell. The Se partial pressure measured at the sample position is fixed at 10$^{-6}$ mbar. In situ Reflection High Energy Electron Diffraction (RHEED) is used to monitor the WSe$_2$ crystal structure during growth. The obtained WSe$_2$ films were then annealed at 1023 K during 15 minutes to improve the crystalline quality. Using this method, centimeter scale (here typically 1$\times$1 cm$^2$) samples can be obtained with a precise control of their thickness, given by the amount of deposited W, Se atoms being in excess by a factor $\sim$20 \cite{Dau2019,Mallet2020,Dosenovic2023}. The graphene on SiC(0001) substrates was slightly doped \cite{Pallecchi2014}. Prior to their introduction in the ARPES chamber, the samples were annealed at 573 K until the pressure stabilised and reached down P$\simeq$10$^{-9}$ mbar (about three hours). The annealing aimed at eliminating most of the contamination adsorbed on the surface.

Prior to the photon energy dependence measurements, we checked the samples to assess their quality and their band structure. Because ARPES is a reciprocal space resolved technique, we can extract qualitative information about the crystallography of the sample like symmetries or surface reconstructions which will manifest in the band structure symmetry and band duplication. Chemical homogeneity can be checked as well by looking at binding energy shifts and sharpness of the bands. We review here the evidences collected by the mean of ARPES on all the samples.

\subsection{2-layer WSe$_2$ sample}
\begin{figure}[!h]
    \centering
    \includegraphics[scale = 1.7]{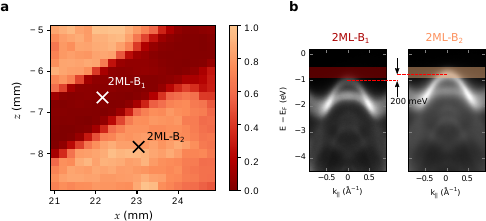}
    \caption{2-layer sample : (a) Intensity map in the range $E_k = -1  \pm 0.2$ eV for sample 2-layer B, $h\nu = 90$ eV - LH polarization. (b) $\Gamma - K$ ARPES cuts on locations representative of the two zones, $h\nu = 60$ eV, LH polarization. The semi-transparent colored areas correspond to the parts of the signal contributing to the intensity map in (a).}
    \label{fig:2MLB_mapXZ}
\end{figure}

This sample was made by MBE on a graphene/SiC(0001) substrate following the procedure described above. Depending on the probed area, the band structure appears to be different. Figure \ref{fig:2MLB_mapXZ}(b) shows two dispersions along $\Gamma-K$ measured on two different locations (labelled $B_1$ and $B_2$) of the sample. There is an obvious binding energy shift in between the two band structures, of the order of 200 meV and the top band is brighter at $B_2$. A X-Y map performed by scanning the beam over a roughly 4$\times$4 mm$^2$ area on the sample surface is presented on Figure \ref{fig:2MLB_mapXZ}(a). The intensity for each pixel is obtained by integrating the intensity over a binding energy range containing the top band on the $B_2$-location (coloured area in \ref{fig:2MLB_mapXZ}(b)). The sample appears to be quite homogeneous at the beam-size scale (around 50$\times$50 µm$^2$), but it is clearly not true at the mm-scale. 

\begin{figure}[!h]
    \centering
    \includegraphics[scale = 1.85]{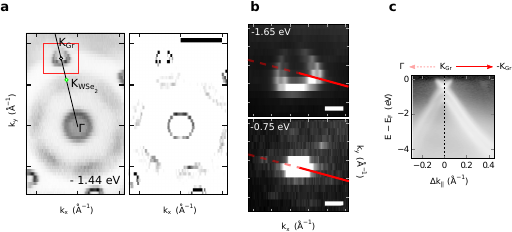}
    \caption{2-layer sample: (a) Constant energy cut of the band structure and second derivative recorded on the B$_1$ zone. The thick black line on top indicates the scale 1 \ang$^{-1}$. (b) Constant energy cuts near the graphene cone at high symmetry point $K_{Gr}$ (area surrounded by the red frame in (a)) at the two indicated binding energies. The thick white line at the bottom indicates the scale 0.2 \ang$^{-1}$. (c) $\Gamma \longrightarrow K_{Gr} \longrightarrow -K_{Gr}$ ARPES cut with logarithmic color scale. The cut is made along the thick red line/dash line highlighted in (b). All measurements are done with $h\nu = 90$ with LH polarization.}
    \label{fig:2MLB_FS}
\end{figure}

\begin{figure}[!h]
    \centering
    \includegraphics[scale = 1]{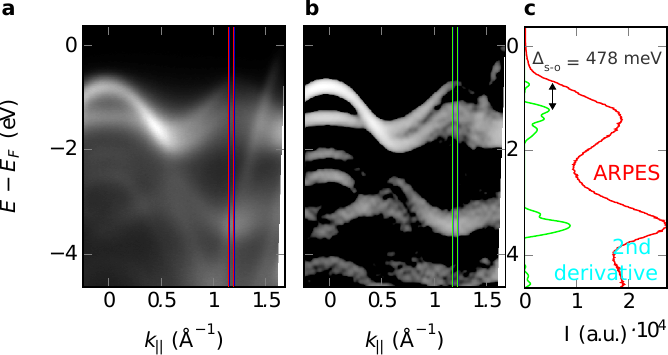}
    \caption{$\Gamma - K$ band dispersion of the 2-layer sample in B$_2$ zone. $h\nu$ = 60 eV, polarization LH. (a) Raw ARPES data, (b) Second derivative, (c) EDC extracted at $K$-point from (a) in red and (b) in green.}
    \label{fig:2ML_ARPES}
\end{figure}

Let's focus on the $B_1$-zone, which presents interesting characteristics. The constant energy cuts shown in Figure \ref{fig:2MLB_FS}(a) a show a relatively well defined band structure with $\Gamma-K$ directions clearly visible for both graphene (denoted $\Gamma-K_{Gr}$ in the figure) and WSe$_2$ ($\Gamma-K_{WSe_2}$). The alignment $\Gamma-K_{WSe_2}-K_{Gr}$ also tells us that the WSe$_2$-layer is truly in epitaxy on graphene. The image does not have excessive azimutal smearing, meaning that the probed WSe$_2$ is essentially single domain. A second derivative of the constant energy surface highlights the double-pocket formed by the two bands at $K_{WSe_2}$ without any other contributions. However, the pocket contour loses some intensity near the $K_{Gr}$ points, certainly because of the large brightness of the latter. A zoom on a $K_{Gr}$ point for two binding energies are presented on Figure \ref{fig:2MLB_FS}(b). The cut at $E_B$=-0.75 eV, near the Dirac point located at $E_D$=-0.32 eV, shows six points forming a hexagon at a distance of 0.4 \ang$^{-1}$ from the center $K_{Gr}$, a signature of a graphene surface reconstruction. The cut at $E_B = -1.65$ eV, shows that the graphene $\pi$-band is actually doubled, meaning that, on the SiC-substrate, coexist single layer (SLG) and bilayer graphene (BLG). This is confirmed by the dispersion displayed in Figure \ref{fig:2MLB_FS}(c).
Looking in detail at Figure \ref{fig:2MLB_mapXZ}(b) we can see that the low-lying band at $\Gamma$ of the $B_1$ location is duplicated. This is because the graphene layer is not completely uniform and two thicknesses of graphene coexist at this location. The magnitude of the charge transfer being dependent on the number of graphene layers \cite{Zhang2021}, the photoemission spectrum of the WSe$_2$ is duplicated. In the $B_2$-zone, there is no trace of splitting in the $\Gamma-K$ dispersion, suggesting a more uniform substrate. We used this zone for the $k_z$-measurements presented in the main text and its band structure in $\Gamma - K$ direction is shown in Figure \ref{fig:2ML_ARPES}.

\clearpage
\subsection{3-layer WSe$_2$ sample}
\begin{figure}[!h]
    \centering
    \includegraphics[scale = 1.7]{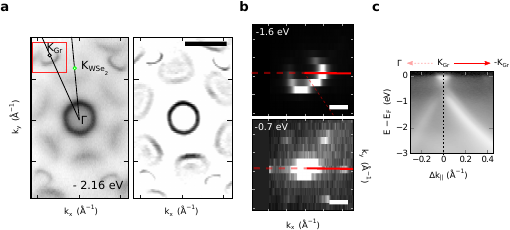}
    \caption{3-layer sample: (a) Constant energy cut of the band structure and second derivative. The thick black line on top indicates the scale 1 \ang$^{-1}$. (b) Detail of the constant energy cut near the graphene cone at high symmetry point $K_{Gr}$ at two energies. The thick white line at the bottom indicates the scale $0.2$ \ang$^{-1}$. (c) $\Gamma \longrightarrow K_{Gr} \longrightarrow -K_{Gr}$ ARPES cut with logarithmic color scale. The cut is made along the thick red line/dash line highlighted in b. All measurements are done with $h\nu = 90$ eV, LH polarization.}
    \label{fig:3ML}
\end{figure}

\begin{figure}[!h]
    \centering
    \includegraphics[scale = 1]{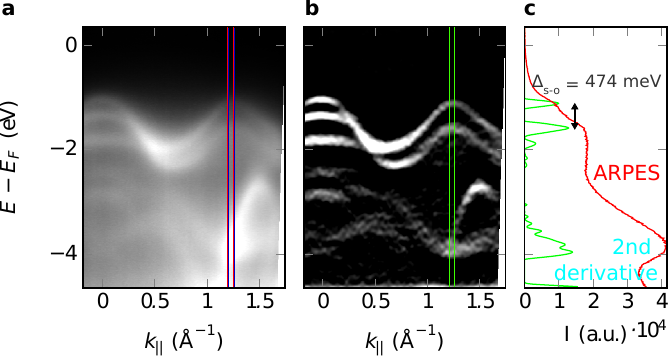}
    \caption{$\Gamma - K$ Band dispersion of the 3-layer sample. $h\nu$=60 eV, LH polarization. (a) Raw ARPES data, (b) Second derivative, (c) EDC extracted at $K$-point from (a) in red and (b) in green.}
    \label{fig:3ML_ARPES}
\end{figure}

The 3-layer WSe$_2$ sample was grown on Mica and then wet-transferred onto a Gr-SiC substrate  \cite{Salazar2022,Dau2019}. The constant energy cuts in Figure \ref{fig:3ML}(a) show a well defined band structure with $K$ points with little azimutal smearing. Both the raw data and their second derivative clearly show the double pocket at $K_{WSe_2}$. Unlike for the 2-layer sample, the $K_{WSe_2}$ and  $K_{Gr}$ points are not aligned but separated by an angle of 13\deg. It is actually not surprising since the sample was grown on a Mica substrate before being transfered onto the graphene layer. The two structures have therefore no reason to be aligned one with respect to the other. The details on the bottom of Figure \ref{fig:3ML}(b) show that the graphene has the same surface reconstruction as in the 2-layer sample, although this time there is only one cone. This is visible both in the zoomed constant energy cut ($E_B$=-1.6 eV) and the ARPES cut in Figure \ref{fig:3ML}(c) implying that the substrate is SLG. A dispersion $\Gamma-K$ recorded at photon energy 60 eV is presented on Figure \ref{fig:3ML_ARPES}. It shows the three expected bands at $\Gamma$ \cite{Nguyen2019}.

\subsection{N-layer WSe$_2$ sample}
\begin{figure}[!h]
    \centering
    \includegraphics[scale = 1.7]{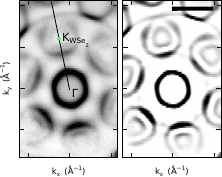}
    \caption{Sample N-ML: Constant energy cut of the band structure (left) and its second derivative (right). The thick black line on top indicates the scale 1 \ang$^{-1}$.}
    \label{fig:NML}
\end{figure}

\begin{figure}[!h]
    \centering
    \includegraphics[scale = 1]{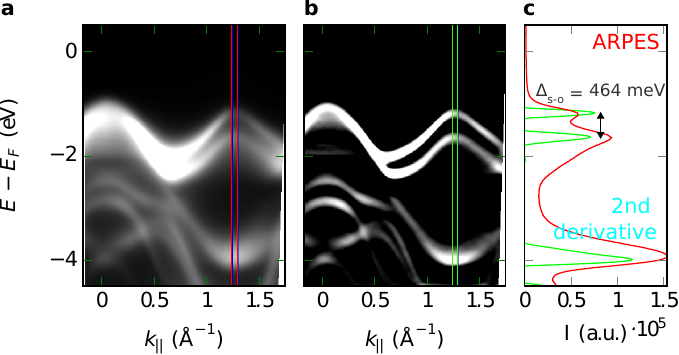}
    \caption{$\Gamma - K$ band dispersion of the N-ML sample. $h\nu$ = 60 eV, polarization LH. (a) Raw ARPES data, (b) Second derivative, (c) EDC extracted at $K$-point from (a) in red and (b) in green.}
    \label{fig:NML_ARPES}
\end{figure}

The last sample used in the work presented here is the N-layer WSe$_2$ (N-ML). The constant energy cut in Figure \ref{fig:NML} show only the band structure of WSe$_2$ film. In this sample, the graphene is not visible anymore because of the high number of WSe$_2$ layers (thick sample). The definition of the ARPES image suggests that the sample is of very high quality with very low azimutal dispersion. Looking closely, it is possible to see ring patterns that hints at some disorder. The overall sharpness of the bands, however, is a strong indicator of the quality of the sample (see Figure \ref{fig:NML_ARPES}).

\newpage
\section{Photon-energy dependence}

\begin{figure}[!h]
    \centering
    \includegraphics[width = 14cm]{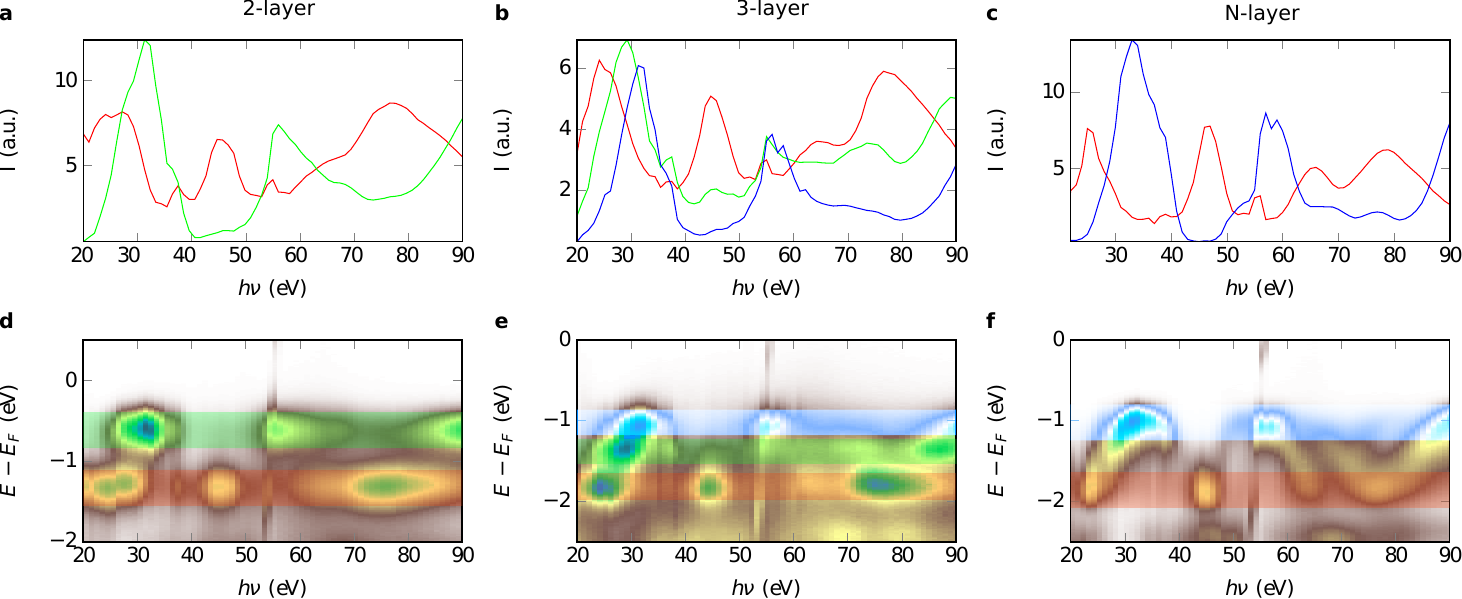}
    \caption{Bottom part: Experimental band intensity variations at $\Gamma$ along the $\Gamma-A$ direction of the reciprocal space for 2 (d), 3 (e) and N (f) layers of WSe$_2$ as a function of the photon energy. Top part: Photoemission intensity integrated over a binding energy range centered on the positions of the bands for those three samples (areas highlighted with colors on the bottom part) (a, b, c).}
    \label{fig:hv_dep}
\end{figure}

Figure \ref{fig:hv_dep} shows the photon-energy dependent photoemission raw signal corresponding to the Figure 3 of the main text prior to the conversion to $k_z$ for the three studied samples. The top part of the figure displays the photoemission intensity integrated over a binding energy range centered on the positions of the bands. It gives a more precise view of their intensity behaviour. 

\section{Tight-binding modelisation of the $k_z$-dispersion}

\subsection{Definition of the model}
To model our experimental data, we developed a tight-binding model on increasingly thick WSe$_2$, retaining only the essential physics of the system and inspired by the derivations presented in references \citenum{amorim2018, kim2016}. $\Gamma$-states are principally composed of one W-5d$_{z^2}$ and two Se-4p$_z$ orbitals. To construct the tight-binding matrix, we use the states $\ket{4p_z^{b(t)}(\mathbf{r}_{b(t),n})}$ with index $b(t)$ corresponding to bottom (top) Se-atoms at position $\mathbf{r}_{b(t),n}$ inside a given layer $n$, as well as the $\ket{5d_{z^2}(\mathbf{r}_{d,n})}$ states of the W-atom located at position $\mathbf{r}_{d,n}$. 
The matrix describing the band structure at $\Gamma$ in a single layer is given by:
\begin{align}
H_{1} = \begin{bmatrix}
e_p & t_{pd} & 0 \\ 
t_{pd} & e_d & t_{pd} \\ 
0 & t_{pd} & e_p\\ 
\end{bmatrix}
\end{align}
and the interlayer Hamiltonian $H_{int}$ is given by:
\begin{align}
H_{int} = \begin{bmatrix}
0 & 0 & 0 & \\ 
0 & 0 & 0 & \\ 
t_{pp} & 0 & 0\\ 
\end{bmatrix}
\end{align}
with on-site energies $e_p$ and $e_d$ and hopping amplitudes $t_{pp} = \bra{p^b_z(\mathbf{r}_{b,n+1})} H_{int} \ket{p^t_z(\mathbf{r}_{t,n})}$ and $t_{pd} = \bra{d_{z^2}(\mathbf{r}_{n,d})} H_1 \ket{p_z^{b(t)}(\mathbf{r}_{b(t),n})}$. We consider that the layers are only coupled through the topmost and bottommost $p_z$ orbitals \cite{Cappelluti2013, kim2016}. These amplitudes are calculated using Slater-Koster integrals $V_{pp\sigma}, V_{pp\pi}, V_{pd\sigma}, V_{pd\pi}$ which, in general, depend on the distance between the considered atomic centers. The expression of $t_{pp}$ and $t_{pd}$ is as follows:
\begin{align}
    t_{pp} &= (V_{pp\sigma} - V_{pp\pi})\left( \dfrac{d_{XX,z}}{d_{XX}}\right )^2 + V_{pp\pi}\\
    t_{pd} &= V_{pd\sigma}\left(\left(\dfrac{d_{MX,z}}{d_{MX}} \right)^2 -\dfrac{1}{2}\left(\dfrac{d_{MX,x}}{d_{MX}} \right)^2\right)\left(\dfrac{d_{MX,z}}{d_{MX}} \right) + \sqrt3\left(\dfrac{d_{MX,z}}{d_{MX}} \right)\left(\dfrac{d_{MX,x}}{d_{MX}} \right)^2V_{pd\pi}
\end{align}
The full Hamiltonian for $N$ layers of WSe$_2$ is a $3N \times 3N$ matrix written as:
\begin{align}
H_{N} = \begin{bmatrix}
H_1 & H_{int} & & &\\ 
H_{int}^T & H_1 & H_{int} & &\\ 
 & H_{int}^T & H_1 & &\\ 
 & & & & \ddots\\
\end{bmatrix}
\end{align}
We directly diagonalize $H_N$ to obtain the new eigenstates of the system, obtaining the $3N$ eigenstates $\ket{\phi_i}$ of energy $E_i$.

\subsection{Calculation of the photoemission current}
The photoemission current is calculated using the Fermi-Golden rule. We assume a damped plane wave final state to mimic partial $k_z$ conservation $ \ket{\mathbf{k}_f- \dfrac{i}{\lambda} \mathbf{e}_{\perp}}$. The photoemission current contributed by the state $\ket{\phi_i}$ = $\sum_{j} P_{ij} \ket{n_j, l_j, m_j, \mathbf{r}_j}$ is:

\begin{align}
   M_{\mathbf{k}_f,\phi_i} &= \bra{\mathbf{k}_f + \dfrac{i}{\lambda} \mathbf{e}_{\perp} } \mathbf{A}\cdot\mathbf{p} \ket{\phi_i}\\
   &= -i\hbar \sum_j P_{ij} \int d^3r \mathbf{A}_0 e^{i\mathbf{k}_{h\nu}\cdot \mathbf{r}} e^{-i\left( \mathbf{k}_{f} + \frac{i}{\lambda} \mathbf{e}_{\perp} \right)\cdot \mathbf{r}} \cdot \nabla R^j_{n_jl_j}(|\mathbf{r}-\mathbf{r}_j|)Y^j_{l_jm_j}(\mathbf{r}-\mathbf{r}_j) \label{M_deriv_1}
\end{align}
Being at low energy ($10 < h\nu < 100$ eV), we neglect the photon momentum $\mathbf{k}_{h\nu}$. We are only interested in the perpendicular direction to the surface i.e. $\mathbf{k}_f = k_z \mathbf{e}_{\perp}$ so that we only keep the component $\mathbf{r}_i \cdot \mathbf{e}_{\perp} = z_i$. Following the derivation of reference \citenum{Moser2017}, equation \eqref{M_deriv_1} becomes:
\begin{align}
   M_{k_z,\phi_i} &\propto \left( -ik_z + \frac{1}{\lambda} \right) \mathbf{A}_0 \cdot \mathbf{e}_{\perp} \sum_j P_{ij} e^{-i\left( k_{z} + \frac{i}{\lambda} \right)\cdot z_i} \int d^3r e^{-i\left( k_{z} + \frac{i}{\lambda} \right) z} R^j_{n_jl_j}(r)Y^j_{l_jm_j}(\mathbf{r})\label{M_deriv_2}
\end{align}
Equation \eqref{M_deriv_2} involves a damped Fourier transform of the orbitals $\ket{nlm}$ that we approximate to the simple Fourier transform of $\braket{\mathbf{k}|nml} = f_{nl}(|\mathbf{k}|)Y_{ml}(\theta_{\mathbf{k}}, \phi_{\mathbf{k}})$ \cite{Moser2017}. In our case, this means $\braket{k_z|nml} = f_{nl}(k_z)Y_{ml}(0, 0)$ with $f_{nl}$ such as:
\begin{align}
    f_{nl}(k) = 4\pi a_*^{3/2}\sqrt{\dfrac{(n-l-1)!}{(n+l)!}}n^2 2^{2l+2} l! \dfrac{(-iy)^l}{(y^2 +1)^{l+2}}C_{n-l-1}^{l+1} \left( \dfrac{y^2 - 1}{y^2 +1} \right)
\end{align}
where $a_* = a_0/Z$ ($a_0$ is the Bohr radius and $Z$ the charge of the nucleus in question) $y = nk/a_*$ and $C_{n-l-1}^{l+1}$ a Gegenbauer polynomial \cite{amorim2018}. It follows that:
\begin{align}
   M_{k_z,\phi_i} &\approx \sum_j P_{ij} e^{-i\left( k_{z} + \frac{i}{\lambda} \right)\cdot z_i} f^j_{n_jl_j}(k_z)Y^j_{l_jm_j}(0,0) \label{M_deriv_3}\\
   &\approx \sum_j  e^{-i\left( k_{z} + \frac{i}{\lambda} \right)\cdot z_i} P_{ij} M_{k_z,j}
\end{align}
Finally we can calculate the total photoemission current as:
\begin{align}
    I_{ph}(k_z, \omega) \approx \sum_i |M_{k_z,\phi_i}|^2 A(\omega - E_i)
\end{align}
where $A(\omega - E_i) = \frac{2\eta}{(\omega - E_i)^2 + \eta^2}$ is the broadened $E_i$ line.

\subsection{Simulation parameters}
\begin{figure}[!h]
    \centering
    \includegraphics[width=6cm]{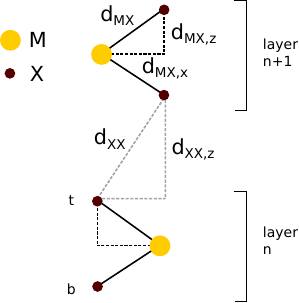}
    \caption{Geometrical model for the 1D $k_z$-dispersion of a MX$_2$ TMD.}
    \label{fig:schematics}
\end{figure}

For the simulation, we used the $V_{pp\sigma} = 1.530$ eV, $V_{pp\pi} = -0.123$ eV, $V_{pd\sigma} = 5.803$ eV, $V_{pd\pi} = -1.081$ eV coefficients from reference \citenum{SilvaGullen2016}. $e_p$ and $e_d = -2.7$ eV are set equal despite the different crystal fields values. This modification accounts for the missing hybridization within the layer. The distances $d_{MX,z} = 3.5881056$ \ang, $d_{MX} = 1.4459472$ \ang, $d_{XX,z} = 4.59299$ \ang, $d_{XX} = 2.55271$ \ang, are determined geometrically (see Figure \ref{fig:schematics}) from the bulk crystallographic parameters \cite{Schutte1987}.
In the specific case of WSe$_2$, one has to encode two contributions for the photoemission signal: one can have the sequence W-Se-Se or the sequence Se-Se-W emitting simultaneously in the $k_z$ direction. This is a simplified version of the screw axis symmetry discussed in reference \citenum{Finteis1997}. This corresponds to a phase factor that we include in the formula of the matrix element:

\begin{align}
   M_{k_z,\phi_i} &\approx \sum_{\varphi \in \{0,2\pi/c\}} \sum_j  e^{-i\left( k_{z} + \varphi + \frac{i}{\lambda} \right)\cdot z_i} P_{ij} M_{k_z,j}
\end{align}
This can lead to additional interferences. For the clarity of exposition, we only keep the phase term $\varphi = 2\pi/c$ which corresponds to the less-suppressed bands in the experimental data.

\clearpage

\section{One-step model ARPES calculations}

The structures of the 1-, 2- and 3-layer as well as bulk WSe$_2$ were constructed using the bulk lattice parameters ($a=3.282$~\r{A}, $c = 12.96$~\r{A}). The electronic structure was calculated using the full potential spin-polarized relativistic Korringa-Kohn-Rostoker method (SPRKKR package)~\cite{Ebert2011}, which solves the Dirac equation using multiple scattering and Green's functions. The 1-, 2-, and 3-layer structures were solved within a repeated slab geometry with vacuum thickness $> 25$~\r{A}. Exchange and correlation effects were treated at the level of local spin density approximation (LSDA) and the basis set was truncated at $l_{\rm max} = 3$. The ARPES calculations were performed in layer-KKR formalism with a semi-infinite surface construction. For the 1-, 2-, and 3-layer structures we set the bulk repeat sequence as vacuum, and therefore the ARPES calculation treats them as truly freestanding thin films. 

\section{Detailed comparison of the models and experiments}

\subsection{Comparison of the three simulation schemes for different 1,2,3,N-layer systems}

\begin{figure}[!h]
    \centering
    \includegraphics[width=16cm]{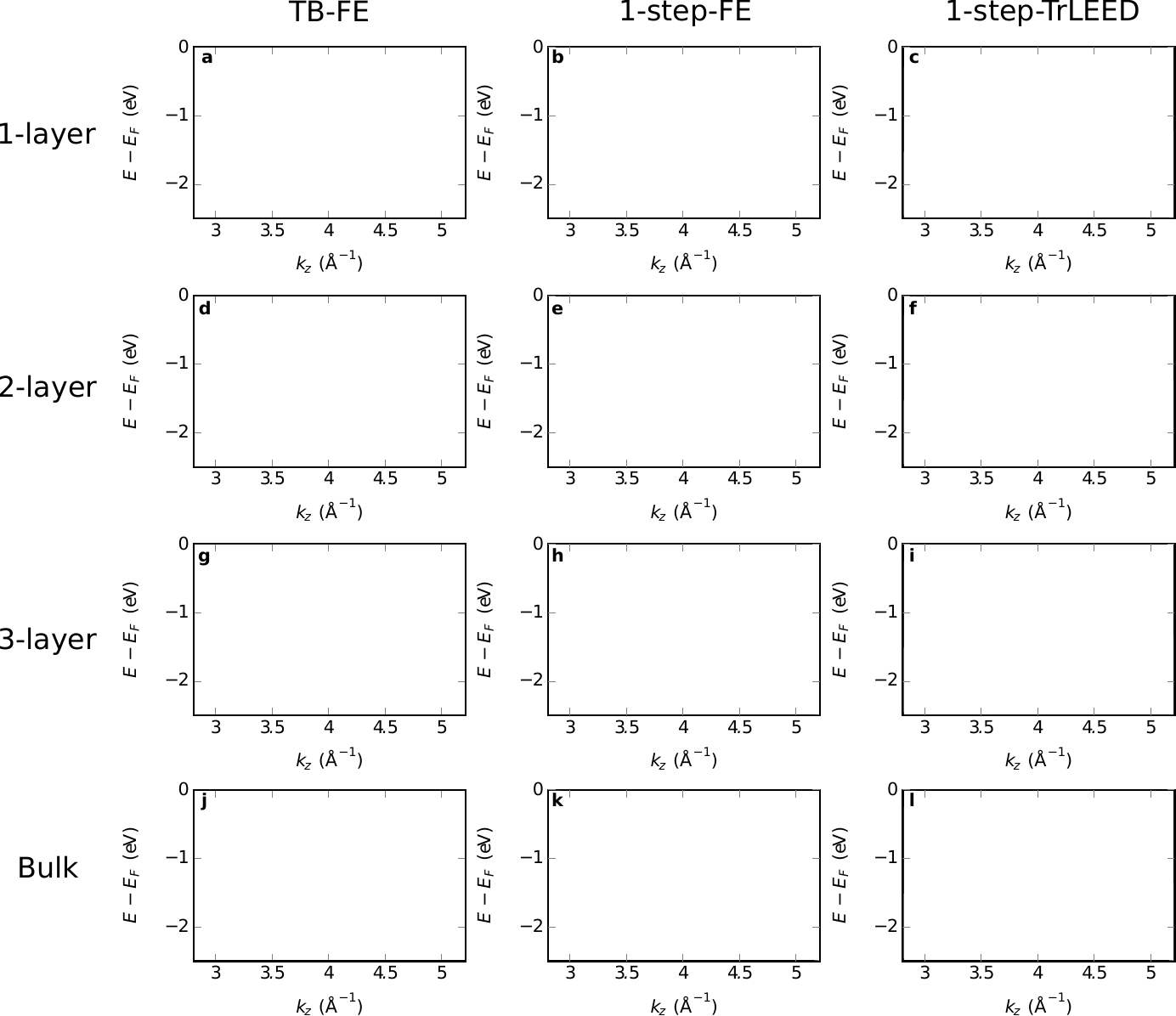}
    \caption{Comparison of $k_z$-dispersions obtained with the different models. Left: TB-FE: tight-binding initial state, free-electron final state. Middle: 1-step-FE: Bloch spectral function initial state, free-electron final state. Right: 1-step-TrLEED: Bloch spectral function initial state, time-reversed LEED final state. These calculations were performed on (from top to bottom) 1-, 2- and 3-layer WSe$_2$ systems as well as on a bulk crystal.}
    \label{fig:comp_simulations}
\end{figure}

Figure \ref{fig:comp_simulations} summarizes the results of the different calculation strategies used in this paper for 1-, 2- and 3-layer WSe$_2$ systems as well as on a bulk crystal.

\subsection{Quantitative comparison of the eigenvalues calculated in the TB and 1-step models with experimental data}

\begin{figure}[!h]
    \centering
    \includegraphics[width=16cm]{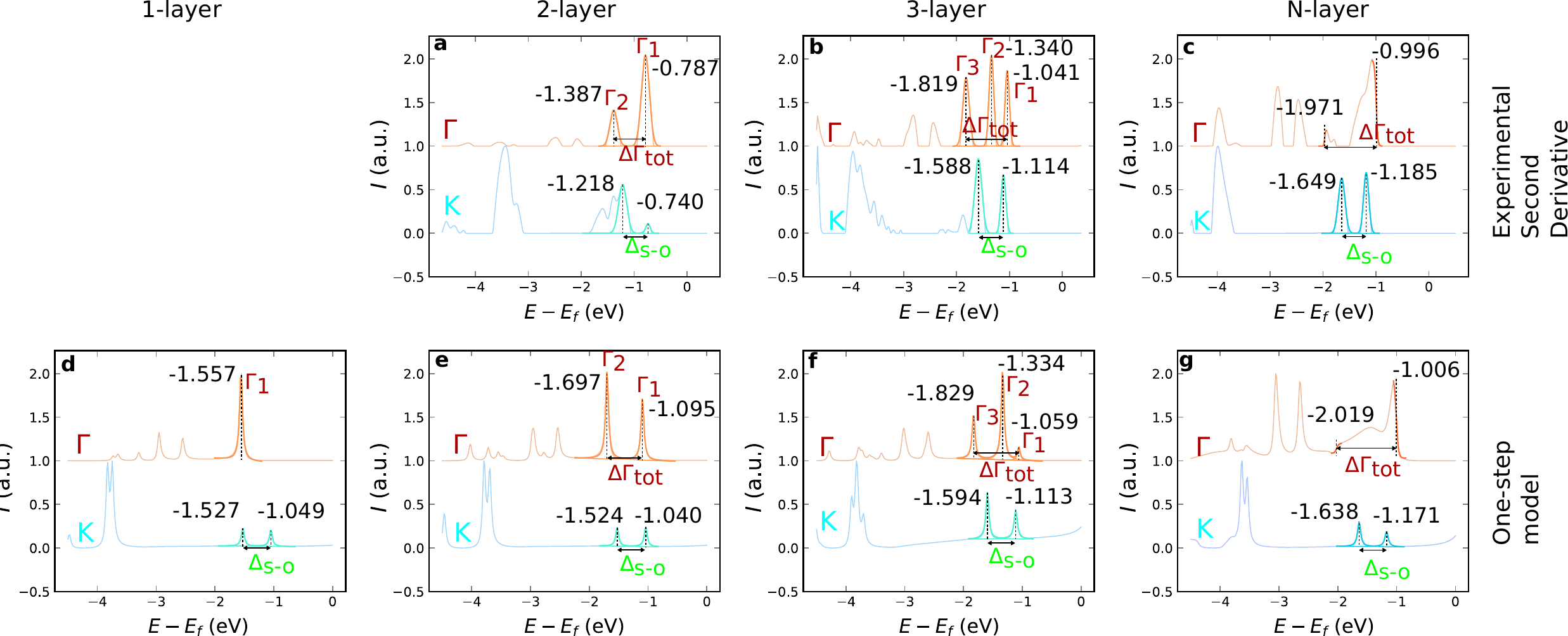}
    \caption{(a-c) Normalised experimental EDC from second derivative signal measured at $h\nu =$ 60 eV for samples 2,3,N-layers (see Figures \ref{fig:2ML_ARPES}, \ref{fig:3ML_ARPES}, \ref{fig:NML_ARPES}). Valence band peaks at $\Gamma$ (labeled $\Gamma_i$) and $K$ are fitted with Gaussian line-shapes. (d-g) Normalised experimental EDC from signal calculated with 1-step-TrLEED model at $h\nu =$ 90 eV for samples 1,2,3,N-layers. Valence band peaks at $\Gamma$ and $K$ are fitted with Lorentzian line-shapes. For the bulk case, $\Delta_{\Gamma_{tot}}$ is obtained by locating the low binding energy side of the band using a fit with a Fermi-Dirac distributions.}
    \label{fig:supp_EDC}
\end{figure}

This section is dedicated to the extraction of the bands binding energies at $\Gamma$ and $K$, both in the experimental data recorded on the different samples and in simulations. This was done by fitting energy dispersion curve (EDC), i.e. vertical cuts in images like the one presented on Figure \ref{fig:NML_ARPES}. To improve the signal to noise ratio, EDCs were averaged over $\pm$0.05 \ang$^{-1}$ around the points of interest. This work was done on the second derivative of experimental images recorded at $h\nu =$ 60 eV, a photon energy which corresponds to a bulk $\Gamma$  point (see Figure \ref{fig:hv_dep}). For the calculations, we used the results obtained with the 1-step-TrLEED model at $h\nu = $ 90 eV, another bulk $\Gamma$  point. Figure \ref{fig:supp_EDC} shows the EDC at $\Gamma$ and $K$ extracted from the experimental data (second derivative) and the calculations. The positions of the peaks are extracted after fitting the data with appropriate line-shapes (Lorentzian for the simulations, Gaussian for the experiments). We are here only interested in the top bands of the valence band, so lower lying bands are excluded from the fit. We calculated different quantities like $\Delta_{\Gamma-K}$, the energy difference between the top of the valence band at $\Gamma$ and at $K$, $\Delta_{s-o}$, the splitting between the two bands at $K$ and $\Delta_{\Gamma-{tot}}$, the total energy width containing all the bands at $\Gamma$ constituting the top of the valence band. The extracted values are summarized in table \ref{tab:gamma_values} and table \ref{tab:K_values}. They compare well to the literature. In the case of 3-layer sample, the reported values differ slightly from the literature value in reference \citenum{Salazar2022} even though the samples are identical. This might indicate a slight deviation from the $\Gamma-K$ cut in the detector or a slight evolution of the sample on a few years time span. 

\begin{table}[!h]
\centering
\begin{tabular}{lllll} 
\hline\hline
                                                                                                                  &                                                          & TB    & 1-step    & EXP    \\ 
\cline{2-5}
 2-layer                  & $\Delta \Gamma_{tot}$ & 0.503 & 0.602 & 0.600  \\ 
\cline{2-5}
                                                                                             \multirow{3}{*}{3-layer} & $\Delta \Gamma_{12}$  & 0.237 & 0.275  & 0.299  \\
                                                                                                                  & $\Delta \Gamma_{23}$  & 0.410 & 0.495  & 0.479  \\
                                                                                                                  & $\Delta \Gamma_{tot}$ & 0.648 & 0.770  & 0.778  \\ 
\cline{2-5}
                                                                                             N-layer                 & $\Delta \Gamma_{tot}$ & 0.737 & 1.013   & 0.975   \\
\hline\hline
\end{tabular}
\caption{Energy differences between the three top bands of the valence band at the $\Gamma$-point ($\Delta\Gamma_{i,j}$ is the energy difference between band $i$ and band $j$; $\Delta\Gamma_{tot}$ is the energy difference between band 1 and band 3) extracted from the TB-FE calculation, the 1-step-based calculations and the experimental data (second derivative). 1-step and experimental values are calculated from the EDCs available in the Supplementary Information. We did not fit 1-step-FE data since they yield the same values as 1-step-TrLEED (same initial state). The values for the TB-FE model are calculated from the eigenvalues after diagonalization of the TB-Hamiltonian.}
\label{tab:gamma_values}
\end{table}

\begin{table}[!h]
\centering
\begin{tabular}{llll} 
\hline\hline
                                                       &                                                          & 1-step    & EXP     \\ 
\cline{2-4}
\multirow{2}{*}{1-layer}  & $\Delta_{\Gamma - K}$ & -0.507 &  (-0.5\cite{Mo2016})       \\
                                                       & $\Delta_{s-o}$                       & 0.477  &   (0.48\cite{Zhang2021})       \\ 
\cline{2-4}
                                 \multirow{2}{*}{2-layer}  & $\Delta_{\Gamma - K}$ & -0.055 & -0.047 (-0.080\cite{Zhang2016})  \\
                                                       & $\Delta_{s-o}$                       & 0.484  & 0.478 (0.489\cite{Zhang2016})   \\ 
\cline{2-4}
                                 \multirow{2}{*}{3-layer}  & $\Delta_{\Gamma - K}$ & 0.054  & 0.073 (0.057\cite{Salazar2022})  \\
                                                       & $\Delta_{s-o}$                       & 0.481  & 0.474 (0.480\cite{Salazar2022})   \\ 
\cline{2-4}
                                 \multirow{2}{*}{N-layer} & $\Delta_{\Gamma - K}$ & 0.165   & 0.189    \\
                                                       & $\Delta_{s-o}$                       & 0.466  & 0.464   \\
\hline\hline
\end{tabular}
\caption{Energy differences (see text for the definitions) between the bands at $\Gamma$ and $K$ extracted from the 1-step calculations and compared to experimental data (second derivative). Calculated from the data in Figure \ref{fig:supp_EDC}. Additional values from the literature are given in parenthesis for comparison with our results.}
\label{tab:K_values}
\end{table}

\clearpage

\bibliography{References.bib}